\documentclass[print,showpacs,pra,10pt,onecolumn,secnumarabic]{revtex4}%
\usepackage{amsfonts}
\usepackage{amsmath}
\usepackage{amssymb}
\usepackage{graphicx}%
\begin{document}
\title{Implementation of many-qubit Grover search by cavity QED}
\author{H. Q. Fan$^{1,2}$}
\email{haoquan.fan@gmail.com (Fax: 011+86-27-8719-9291)}
\author{W. L. Yang$^{1,2}$}
\email{yanglingfeng1980@yahoo.com.cn}
\author{X. R. Huang$^{1}$}
\email{hxueren@wipm.ac.cn}
\author{M. Feng$^{1}$}
\email{mangfeng1968@yahoo.com}
\affiliation{$^{1}$State Key Laboratory of Magnetic Resonance and Atomic and Molecular
Physics, Wuhan Institute of Physics and Mathematics, Chinese Academy of
Sciences, Wuhan 430071, China}
\affiliation{$^{2}$Graduate School of the Chinese Academy of Sciences, Beijing 100049, China}

\begin{abstract}
We explore the possibility of $N$-qubit $(N>3)$ Grover search in cavity QED,
based on a fast operation of $N$-qubit controlled phase-flip with atoms in
resonance with the cavity mode. We demonstrate both analytically and
numerically that, our scheme could be achieved efficiently to find a marked
state with high fidelity and high success probability. As cavity decay is
involved in our quantum trajectory treatment, we could analytically understand
the implementation of a Grover search subject to dissipation, which would be
very helpful for relevant experiments.

\end{abstract}

\maketitle

\section{introduction}

Cavity quantum electrodynamics (QED) has got much progress over past several
years \cite{review}. People have achieved various optical and microwave
cavities, with which some remarkable quantum phenomena have been demonstrated,
such as strong coupling of a single atom with a single photon \cite{infer},
quantum logic gating between two atomic qubits \cite{haro1}, photonic blocking
\cite{photon}, efficient generation of single photons \cite{generation}, and
so on.

In the present work, we will investigate how to realize a Grover search with
more than three qubits by cavity QED. The Grover search, as one of the most
frequently mentioned quantum algorithms, could effectively exemplifies the
potential speed-up offered by quantum computers \cite{grover}. Recently, many
authors have dedicated themselves to the search algorithm by adiabatic
evolution methods \cite{daems,roland,sch} or nonadiabatic scheme \cite{perez}.
Although achievement of these schemes needs stringent conditions and demanding
techniques \cite{tong}, they are really wonderful ideas. On the other hand,
some authors had addressed the effect of decoherence \cite{azuma}, noise
\cite{pablo}, and gate imperfection or errors \cite{long} on the efficiency of
quantum algorithms. We also noticed that, there had been intensive interests
in achieving Grover search algorithm theoretically and experimentally by using
NMR systems \cite{jones}, trapped ions \cite{feng,fuji,brick}, linear optical
elements \cite{walther,dodd}, cavity\ QED \cite{osnaghi,yama,yang,deng,joshi},
and superconducting meso-circuits \cite{naka}.

Our interest is in the experimental feasibility of Grover search by a
microwave cavity. A recent work \cite{yang}, with two of the authors joined
in, showed a three-qubit Grover search implementation in a microwave cavity.
As it makes use of the resonant interaction of the atoms with the cavity mode
and consideres the detrimental influence from the cavity decay, the
implementation of \cite{yang} is close to the reach of current technique.
However, that scheme is hard to be extended to the case with more than three
qubits due to the method \cite{chen} used there. In the present work, we will
show that, after some modification of the method in \cite{chen}, we could
accomplish a high-fidelity conditional phase flip (CPF) for many atomic qubits
flying through a microwave cavity, based on which an $N$-qubit $(N>3)$ Grover
search might be achieved in a straightforward way. The main idea is that, we
make a smart qubit encoding on the atomic levels, and send the atoms through a
microwave cavity for resonant interaction. As long as we could exactly control
the interaction time and the trajectories of the flying atoms, we may
accomplish the $N$-qubit CPF gate by one step of implementation, based on
which we could efficiently carry out an $N$-qubit Grover search with high
fidelity. However, as the Grover search with more than two qubits works only
probabilistically and also as our one-step implementation of many-qubit CPF
gate intrinsically owns imperfection, the situation in our case is much more
complicated than in two-qubit case. To make our scheme more realistic, we
involve the cavity decay in our treatment, and as done in \cite{yang}, we will
try to present some analytical expressions for the state evolution during our
implementation. As a result, we could explicitly assess how well a Grover
search is made subject to these disadvantageous factors.

\section{N-qubit CPF gate subject to cavity decay}

As described in \cite{yang,chen}, we consider the resonant interaction of $N$
three-level atoms with a single-mode cavity. As shown in Fig. 1, the three
levels can be described by $\left\vert i_{j}\right\rangle $, $\left\vert
g_{j}\right\rangle $, and $\left\vert e_{j}\right\rangle $, in which the
subscript $j$ means the $j$th atom, the states $\left\vert e_{j}\right\rangle
$ are excited states and $\left\vert i_{j}\right\rangle $, $\left\vert
g_{j}\right\rangle $ are ground states. As $\left\vert g_{j}\right\rangle $
and $\left\vert e_{j}\right\rangle $ are coupled in resonance by the cavity
mode, the states\ $\left\vert i_{j}\right\rangle $ are not involved in the
interaction with the cavity mode throughout our treatment.\ We encode the
first qubit $\left\vert 0\right\rangle _{1}$ $(\left\vert 1\right\rangle
_{1})$ on $\left\vert e_{1}\right\rangle $ ($\left\vert g_{1}\right\rangle ),$
and other qubits $\left\vert 0\right\rangle _{k}$ $(\left\vert 1\right\rangle
_{k})$ on $\left\vert i_{k}\right\rangle $ $(\left\vert g_{k}\right\rangle )$
with $k=2,3,...,N$. Considering the cavity with weak decay, which implies no
photon actually leaking out of the cavity during our implementation, we may
write the Hamiltonian in units of $\hbar=1$ as,%
\begin{equation}
H=\sum_{i=1}^{N}\Omega_{i}\left(  a^{\dagger}\sigma_{i}^{-}+a\sigma_{i}%
^{+}\right)  -i\frac{\kappa}{2}a^{\dagger}a,
\end{equation}
where $\Omega_{i}$ is the coupling constant of the $i$th atom to the cavity
mode, and the atomic spin operators for raising and lowering are $\sigma
_{i}^{+}=\left\vert e_{i}\right\rangle \left\langle g_{i}\right\vert $ and
$\sigma_{i}^{-}=\left\vert g_{i}\right\rangle \left\langle e_{i}\right\vert ,$
respectively. $a^{\dagger}$ ($a$) is the creation (annihilation) operator of
the cavity mode.\ We will first consider an implementation of $N$-qubit CPF
under cavity dissipation, based on which an $N$-qubit Grover search would be
achieved in the next section.

Assuming the initial atomic state in $\left\vert e_{1}\right\rangle
\prod_{l,j=2,l\neq j}^{N}\left\vert g_{l}\right\rangle \left\vert
i_{j}\right\rangle $ and the cavity initially in the vacuum state $\left\vert
0\right\rangle $, we could deduce the evolved state by directly solving the
Schr\H{o}dinger equation regarding Eq. (1), which yields \cite{chen},%
\begin{align}
\left\vert e_{1}\right\rangle \prod_{l,j=2,l\neq j}^{N}\left\vert
g_{l}\right\rangle \left\vert i_{j}\right\rangle \left\vert 0\right\rangle  &
\longrightarrow\left[  \frac{\Omega_{1}^{2}}{G_{m}^{2}}\exp\left(  -\kappa
t/4\right)  \left(  \cos\left(  A_{m+1,\kappa}t\right)  +\frac{\kappa
}{4A_{m+1,\kappa}}\sin\left(  A_{m+1,\kappa}t\right)  \right)  +\frac
{G_{m}^{2}-\Omega_{1}^{2}}{G_{m}^{2}}\right]  \times\nonumber\\
&  \left\vert e_{1}\right\rangle \prod_{l,j=2,l\neq j}^{N}\left\vert
g_{l}\right\rangle \left\vert i_{j}\right\rangle \left\vert 0\right\rangle
+\frac{\Omega_{1}^{2}}{G_{m}^{2}}\left[  \exp\left(  -\kappa t/4\right)
\left(  \cos\left(  A_{m+1,\kappa}t\right)  +\frac{\kappa}{4A_{m+1,\kappa}%
}\sin\left(  A_{m+1,\kappa}t\right)  \right)  -1\right]  \times\nonumber\\
&  \left\vert g_{1}\right\rangle \sum_{k=2}^{m+1}\Omega_{k}\left\vert
e_{k}\right\rangle \prod_{l,j=2,l\neq j\neq k}^{N}\left\vert g_{l}%
\right\rangle \left\vert i_{j}\right\rangle \left\vert 0\right\rangle -\left(
i\Omega_{1}/A_{m+1,\kappa}\right)  \sin\left(  A_{m+1,\kappa}t\right)
\left\vert g_{1}\right\rangle \prod_{l,j=2,l\neq j}^{N}\left\vert
g_{l}\right\rangle \left\vert i_{j}\right\rangle \left\vert 1\right\rangle ,
\end{align}
\bigskip where $G_{m}=\sqrt{\sum_{j=1}^{m+1}\Omega_{j}^{2}}$ and
$A_{m+1,\kappa}=\sqrt{\sum_{j=1}^{m+1}\Omega_{j}^{2}-\kappa^{2}/16}$, with
$m$\ the number of the atoms initially in the ground state $\left\vert
g\right\rangle $.

In order to accomplish an $N$-qubit CPF gate, we should also pay attention to
the evolution with the initial states $\left\vert e_{1}\right\rangle
\prod_{j=2}^{N}\left\vert i_{j}\right\rangle \left\vert 0\right\rangle $,
which is%
\begin{align}
\left\vert e_{1}\right\rangle \prod_{j=2}^{N}\left\vert i_{j}\right\rangle
\left\vert 0\right\rangle  &  \longrightarrow\left[  \exp\left(  -\kappa
t/4\right)  \left(  \cos\left(  A_{1,\kappa}t\right)  +\frac{\kappa
}{4A_{1,\kappa}}\sin\left(  A_{1,\kappa}t\right)  \right)  \right]  \times\\
&  \left\vert e_{1}\right\rangle \prod_{j=2}^{N}\left\vert i_{j}\right\rangle
\left\vert 0\right\rangle -\left(  i\Omega_{1}/A_{1,\kappa}\right)
\sin\left(  A_{1,\kappa}t\right)  \left\vert g_{1}\right\rangle \prod
_{j=2}^{n}\left\vert i_{j}\right\rangle \left\vert 1\right\rangle ,\nonumber
\end{align}
with $A_{1,\kappa}=\sqrt{\Omega_{1}^{2}-\kappa^{2}/16}$. Assuming the coupling
constant satisfies the condition $\Omega=\Omega_{2}=\Omega_{3}=\cdot\cdot
\cdot=\Omega_{N}\gg\Omega_{1},$ we have, under the approximation of short time
evolution with $\sin\left(  A_{1,\kappa}t\right)  <<\cos\left(  A_{1,\kappa
}t\right)  ,$ Eq. (3) reduced to%
\begin{equation}
\left\vert e_{1}\right\rangle \prod_{j=2}^{N}\left\vert i_{j}\right\rangle
\left\vert 0\right\rangle \longrightarrow\alpha\left\vert e_{1}\right\rangle
\prod_{j=2}^{N}\left\vert i_{j}\right\rangle \left\vert 0\right\rangle ,
\end{equation}
with $\alpha=\left[  \exp\left(  -\kappa t/4\right)  \left(  \cos\left(
A_{1\kappa}t\right)  +\frac{\kappa}{4A_{1\kappa}}\sin\left(  A_{1\kappa
}t\right)  \right)  \right]  $. Under the same condition as above, Eq. (2)
similarly reduces to%
\begin{equation}
\left\vert e_{1}\right\rangle \prod_{\substack{j,k=2,\\j\neq k}}^{N}\left\vert
g_{k}\right\rangle \left\vert i_{j}\right\rangle \left\vert 0\right\rangle
\longrightarrow\beta_{m}\left\vert e_{1}\right\rangle \prod
_{\substack{j,k=2,\\j\neq k}}^{N}\left\vert g_{k}\right\rangle \left\vert
i_{j}\right\rangle \left\vert 0\right\rangle ,
\end{equation}
with $\beta_{m}=\left[  \frac{\Omega_{1}^{2}}{G_{m}^{2}}\exp\left(  -\kappa
t/4\right)  \left(  \cos\left(  A_{(m+1)\kappa}t\right)  +\frac{\kappa
}{4A_{(m+1)\kappa}}\sin\left(  A_{(m+1)\kappa}t\right)  \right)  +\frac
{G_{m}^{2}-\Omega_{1}^{2}}{G_{m}^{2}}\right]  $. In order to make a CPF
gating, we need $\alpha\rightarrow-1$ in Eq. (4), which corresponds to
$t=\pi/\sqrt{\Omega_{1}^{2}-\kappa^{2}/16}$. As a result, the coefficient
$\beta_{m}$ becomes%
\begin{equation}
\beta_{m}=\frac{-\alpha}{1+m\eta^{2}}(\cos\vartheta+\mu\sin\vartheta
/\sqrt{16+16m\eta^{2}-\mu^{2}})+\frac{m\eta^{2}}{1+m\eta^{2}},
\end{equation}
where $\vartheta=\pi\sqrt{1+16m\eta^{2}/(16-\mu^{2})},$ $\ \eta=\Omega
/\Omega_{1}$ and $\alpha=-\exp\left(  -\pi\mu/\sqrt{16-\mu^{2}}\right)
\ $with $\mu=\kappa/\Omega_{1}$. It is easily verified that $\alpha
\rightarrow-1$ and $\beta_{m}\rightarrow1$ in the case of $\kappa
\rightarrow0.$ We may take $N=4$ as an example to show the CPF more
explicitly. Following the deduction above, we have the four-qubit CPF written
as $U^{\left(  4\right)  }=J_{0000}=diag\left\{  \alpha,\beta_{1},\beta
_{1},\beta_{1},\beta_{2},\beta_{2},\beta_{2},\beta_{3}%
,1,1,1,1,1,1,1,1\right\}  $ with respect to the bases $\left\vert e_{1}%
i_{2}i_{3}i_{4}\right\rangle $, $\left\vert e_{1}i_{2}i_{3}g_{4}\right\rangle
$, $\left\vert e_{1}i_{2}g_{3}i_{4}\right\rangle $, $\left\vert e_{1}%
g_{2}i_{3}i_{4}\right\rangle $, $\left\vert e_{1}i_{2}g_{3}g_{4}\right\rangle
$, $\left\vert e_{1}g_{2}i_{3}g_{4}\right\rangle $, $\left\vert e_{1}%
g_{2}g_{3}i_{4}\right\rangle $, $\left\vert e_{1}g_{2}g_{3}g_{4}\right\rangle
$, $\left\vert g_{1}i_{2}i_{3}i_{4}\right\rangle $, $\left\vert g_{1}%
i_{2}i_{3}g_{4}\right\rangle $, $\left\vert g_{1}i_{2}g_{3}i_{4}\right\rangle
$, $\left\vert g_{1}g_{2}i_{3}i_{4}\right\rangle $, $\left\vert g_{1}%
i_{2}g_{3}g_{4}\right\rangle $, $\left\vert g_{1}g_{2}i_{3}g_{4}\right\rangle
$, $\left\vert g_{1}g_{2}g_{3}i_{4}\right\rangle $, and $\left\vert g_{1}%
g_{2}g_{3}g_{4}\right\rangle ,$ respectively, where the latter eight states do
not evolve under the Hamiltonian of Eq. (1) in the case of vacuum cavity
state. To assess the CPF gating, we assume the atoms to be initially in
$\left\vert \Psi_{0}\right\rangle =\frac{1}{\sqrt{2}}\left(  \left\vert
g_{1}\right\rangle +\left\vert e_{1}\right\rangle \right)  \left(  \left\vert
g_{2}\right\rangle +\left\vert i_{2}\right\rangle \right)  \bullet
\bullet\bullet\left(  \left\vert g_{N}\right\rangle +\left\vert i_{N}%
\right\rangle \right)  $. We may calculate the fidelity and the success
probability by employing the relations $F=\overline{\left\langle \Psi
_{f}\right\vert U^{\left(  n\right)  \dagger}\left\vert \Psi_{0}\right\rangle
\left\langle \Psi_{0}\right\vert U^{\left(  n\right)  }\left\vert \Psi
_{f}\right\rangle }$ \cite{Poya}\ and $P=\overline{\left\langle \Psi_{f}%
|\Psi_{f}\right\rangle }$ \cite{yang,chen} As shown in Fig. 2, we could reach
considerably high fidelity $F$ and high probability of success $P$ as long as
$\eta$ $\longrightarrow10,$ even in the case of relatively large decay rate.
In what follows, we will employ our CPF to carry out Grover search under weak dissipation.

\section{N-qubit Grover search}

Generally speaking, provided that initial state of the system has been
prepared in an average superposition state $\left\vert \Psi_{0}\right\rangle
=(1/\sqrt{2^{N}})\sum\limits_{i=0}^{2^{N}}\left\vert i\right\rangle ,$ Grover
search can be depicted as the iterative operation $\widehat{D}^{(N)}J_{\tau}$
by at least $\pi\sqrt{2^{N}}/4$ times for finding the marked state $\left\vert
\tau\right\rangle $ with an optimal probability$,$ where the quantum phase
gate $J_{\tau}=I-2\left\vert \tau\right\rangle \left\langle \tau\right\vert $
(with $I$ being the identity matrix) plays an important role to invert the
amplitude of the marked state, and the diffusion transform $\widehat{D}^{(N)}$
is defined as $\widehat{D}_{ij}=2/K-\delta_{ij}$ $(i,j=1,2,...,K,$ $K=2^{N}).$
Considering the qubit subspace spanned by \{$\left\vert e_{1}\right\rangle $,
$\left\vert g_{1}\right\rangle $, $\left\vert i_{2}\right\rangle $,
$\left\vert g_{2}\right\rangle $ ... $\left\vert i_{N-1}\right\rangle $,
$\left\vert g_{N-1}\right\rangle $, $\left\vert i_{N}\right\rangle $,
$\left\vert g_{N}\right\rangle $\}, to make a Grover search, we could
construct following transformation,%
\begin{equation}
\allowbreak Q^{(N)}=\allowbreak W^{\otimes N}J_{00\cdots0}^{(N)}\allowbreak
W^{\otimes N}J_{\rho}=\allowbreak W^{\otimes N}\allowbreak J_{e_{1}i_{2}\cdots
i_{n}}^{(N)}W^{\otimes N}J_{\rho}=-\hat{D}^{(N)}J_{\rho},
\end{equation}
where $J_{00\cdots0}^{(N)}=diag\{-1,1,\cdots,1\}=I^{(N)}-2\left\vert
00\cdots0\right\rangle \left\langle 00\cdots0\right\vert $ (with superscript
$(N)$ denoting $N$-qubit case and $\cdots$ denoting $N$ elements) and the
Hadamard gate in our computational subspace is given by%
\begin{equation}
W^{\otimes N}=\prod\limits_{i=1}^{N}W_{i}=\left(  \frac{1}{\sqrt{2}}\right)
^{N}%
\begin{bmatrix}
1 & 1\\
1 & -1
\end{bmatrix}
\otimes%
\begin{bmatrix}
1 & 1\\
1 & -1
\end{bmatrix}
\otimes\cdots\otimes%
\begin{bmatrix}
1 & 1\\
1 & -1
\end{bmatrix}
. \label{21}%
\end{equation}
In the case of $N$ qubits, Eq. (7) implies that the diffusion transform
$\hat{D}^{(N)}=-\allowbreak W^{\otimes N}\allowbreak J_{e_{1}i_{2}\cdots
i_{n}}^{(N)}W^{\otimes N}$ is always unchanged no matter which state is to be
searched. The only change is the operator $J_{\rho}$ for different marked
states. Based on the CPF gate constructed in last section, we have
$J_{00\cdots0}^{(N)}\approx diag\{-1,1,\cdots,1\}=U_{CPF}^{(N)},$ from which
we could achieve the gate $J_{11\cdots1}^{(N)}$ as,%
\begin{equation}
J_{11\cdots1}^{(N)}=S_{x,N}S_{x,N-1}\cdots\sigma_{x,1}J_{00\cdots0}%
^{(N)}\sigma_{x,1}\cdots S_{x,n-1}S_{x,n}, \label{22}%
\end{equation}
where $S_{x,j}=|i_{j}\rangle\langle g_{j}|+|g_{j}\rangle\langle i_{j}|$ with
$j\neq1,$ and $\sigma_{x,1}=|e_{1}\rangle\langle g_{1}|+|g_{1}\rangle\langle
e_{1}|.$ To achieve $\allowbreak Q^{(N)}$, we will construct the CPF gate
$J_{\rho}=I-2\left\vert \rho\right\rangle \left\langle \rho\right\vert .$ For
example, in the case of $N=10$, the number of the involved quantum states is
$\allowbreak2^{10}$, and the operation is to label a marked state by
$\allowbreak J_{\rho}$ with $\rho$ one of the states from $\{\left\vert
0000000000\right\rangle ,\left\vert 0000000001\right\rangle ,\left\vert
0000000010\right\rangle ,\cdots,\left\vert 1111111111\right\rangle \}.$ To
carry out the ten-qubit Grover search, we need two ten-qubit Hadamard gates
$W^{\otimes10}$. Based on the CPF gate marking the state $\left\vert
g_{1}g_{2}g_{3}g_{4}g_{5}g_{6}g_{7}g_{8}g_{9}g_{10}\right\rangle ,$ we could
construct other $2^{10}-1$ gates for the marking job by single-qubit
rotations. For example,%
\begin{align}
J_{g_{1}g_{2}g_{3}g_{4}g_{5}g_{6}g_{7}g_{8}g_{9}g_{10}}  &  =J_{1111111111}%
=\nonumber\\
&  S_{x,10}S_{x,9}S_{x,8}S_{x,7}S_{x,6}S_{x,5}S_{x,4}S_{x,3}S_{x,2}%
\sigma_{x,1}J_{0000000000}\sigma_{x,1}S_{x,2}S_{x,3}S_{x,4}S_{x,5}%
S_{x,6}S_{x,7}S_{x,8}S_{x,9}S_{x,10}.
\end{align}
So with a state marked, and with the ten-qubit diffusion transform $\hat
{D}^{(10)}$\ which is generated by combining two Hadamard gates $W^{\otimes
10}$ with $J_{0000000000}$, a full Grover search for ten qubits is available.

Let us consider a ten-qubit Grover search for the marked state $\left\vert
e_{1}i_{2}g_{3}g_{4}i_{5}i_{6}i_{7}i_{8}i_{9}i_{10}\right\rangle $ as an
example. The Rydberg atoms from the 1st to the 10th are input, as shown in
Fig. 1, with principal quantum numbers $49$, $50$ and $51$ denoting
$|g_{j}\rangle$, $|i_{j}\rangle$ and $|e_{j}\rangle,$ respectively. Suppose
the atoms initially prepared in a superposition $\left\vert \Psi
_{0}\right\rangle =2^{-5}\times(\left\vert e_{1}\right\rangle +\left\vert
g_{1}\right\rangle )(\left\vert i_{2}\right\rangle +\left\vert g_{2}%
\right\rangle )(\left\vert i_{3}\right\rangle +\left\vert g_{3}\right\rangle
)\cdot\cdot\cdot(\left\vert i_{10}\right\rangle +\left\vert g_{10}%
\right\rangle )$. We plot in Fig. 3 for the success rate of the Grover search
in different cases. As the Grover search involving more than two qubits is
intrinsically probabilistic, how to obtain an optimal search is a problem of
interest. In contrast to the previous study for an ideal Grover search
\cite{boyer}, our analytical expressions along with some numerics could
identify the optimal search under cavity decay and other imperfection due to
gating. We know from Fig. 3 that, because of the cavity decay, the more
iteration steps, the more detrimental effect from the cavity decay involved.
As a result, we prefer to have less iteration steps for an optimal search. The
detrimental effect from the cavity decay is also reflected in the figure that
the maximum value of P is slightly lower in the case of larger decay rate.

\section{discussion}

Our CPF gating plays the essential role in our Grover search scheme, which
much reduces the implementation time compared to that by a series of two-qubit
conditional gates and single-qubit gates \cite{decompo}. Sec II has shown that
the CPF gating time only depends on the weakest coupling $\Omega_{1}$ and the
cavity decay rate $\kappa,$ but irrelevant to the qubit number. So our scheme
could keep constant implementation time no matter how many atoms are involved,
which favors a large-scale QIP. As a result, if we suppose the single-qubit
rotation takes negligible time compared to the CPF gating, then our Grover
search could be carried out by a constant time irrelevant to the qubit number.
Specifically, assuming $\Omega=2\pi\times49$ kHz \cite{review}$,$ $\Omega
_{1}=2\pi\times4.9$ kHz$,$ and $\kappa=0.1\Omega_{1},$ we have $t_{0}%
=\pi/\sqrt{\Omega_{1}^{2}-\kappa^{2}/16}\approx102$ $\mu s$, which is much
shorter than either the cavity decay time, i.e., $2\pi/\kappa\approx2$ ms, or
the Rydberg atomic lifetime $30$ ms \cite{review}. However, for a Grover
search, the operation $Q^{(N)}$ has to be made for several times depending on
the specific search task. In the case of $N=10,$ the total required time for a
Grover search would be about $2[\pi\sqrt{2^{N}}/4]\times$ $t_{0}\approx$ $5$
ms which is comparable to the cavity decay time.

As shown in Fig.3, our numerics shows that the iteration steps for optimal
search are $25$, $23$, $17,$ respectively, in the case of $\mu=0,$ $0.05,$ and
$0.1$. Since the operation time $t_{0}=\pi/\sqrt{\Omega_{1}^{2}-\kappa^{2}%
/16}$ varies very little in the case of weak cavity dissipation with respect
to the ideal case, the less iteration steps under dissipation would definitely
lead to reduction of our implementation time. Straightforward calculation
shows that the required time for an optimal search are\ $5$ ms, $\ 4.6$ ms,
and $3.4$ ms, respectively, in the case of $\mu=0,$ $0.05,$ and $0.1$. It is
of importance to have a fast implementation of Grover search in view of
decoherence. But if we seriously consider the values in above calculation, we
could find that $\mu=0.05,$ and $0.1$ correspond to dissipation time of 4 ms
and 2 ms, respectively. As a result, to have a fast implementation with our
scheme, we must increase the coupling strengths $\Omega$ and $\Omega_{1}$ by
at least ten times$,$ which makes sure the required implementation time to be
much shorter than the dissipation time.

In current microwave cavity experiments \cite{review}, we may consider the
interaction between atoms and the cavity mode by $\Omega\cos(2\pi
z/\lambda_{0})\exp(-r^{2}/w^{2})\sim\Omega\cos(2\pi z/\lambda_{0})$
\cite{review}, where $\Omega$ is the coupling strength at the cavity center,
$r$ is the distance of the atom away from the cavity center, and $\lambda_{0}$
and $w$ are the wavelength and the waist of the cavity mode, respectively. To
meet the condition of $\Omega=10\Omega_{1}$,\ we may send the first atom going
through the cavity along y axis deviating from the nodes by $(\lambda_{0}%
/2\pi)\arccos(1/10)$, but other atoms through the antinodes. Specifically, for
the N input atoms (suppose N being an even number for simplicity), we should
have the tracks of the atomic movement as $z_{1}=-(N/2)\lambda_{0}%
+(\lambda_{0}/2\pi)\arccos(1/10)$, $z_{2}=-(N/2-1)\lambda_{0},$ ....,
$z_{N}=(N/2)\lambda_{0}$. Moreover, with current cavity QED techniques, the
design in Fig. 1 could be realized by 50 separate microwave cavities with each
Ramsey zone located by a cavity. Since each microwave cavity is employed to
carry out a ten-qubit phase gate $J_{00\cdots0}^{(10)}$, the cavities should
be identical. While searching for different states, we employ different
single-qubit operations. So the Ramsey zones should be long enough to finish
the necessary single-qubit operations.

However, it is highly challenging with current experimental technology to
simultaneously send many Rydberg atoms through a cavity with precise tracks
and velocities although two Rydberg atoms going through a microwave cavity
simultaneously have been achieved \cite{osnaghi}. Here we assess the
infidelity due to the imperfection in atomic velocity by a simple example
below. Consider the simple situation that the first atom moves a little bit
slower in the cavity than other atoms, that is, $t_{1}=t_{0}+\delta t$,
$t_{2}=t_{3}=\cdot\cdot\cdot=t_{n}=t_{0}$, with $t_{0}=\pi/\sqrt{g_{1}%
^{2}-\kappa^{2}/16}$. As shown in Fig.4 with $\eta=10$, we can see that the
infidelity increases with both $\delta t$ and $\mu$. In actual experiment, the
situation is much more complicated. Besides the diversity of atomic velocity,
there will be other imperfections such as classical pulse imperfection,
differences of the cavities, deflected atomic trajectories and so on. Our
simple estimate could show the importance of highly controllable movement of
the atoms in implementing our scheme.

\section{conclusion and acknowledgements}

In conclusion, we have studied many-qubit CPF gate in cavity QED under weak
cavity dissipation, and explored the possibility of many-qubit Grover search
in a dissipative cavity QED system. As a theoretical work, like previous
publication in cavity QED, e.g., \cite{hua}, we have simplified the complexity
in the realistic system, and tried to present some analytical expressions for
the state evolution subject to the cavity decay. Although we took an example
involving ten qubits, our approach could also be easily used for three or four
atomic qubits in cavity QED, which might be closer to experimental
realization. On the other hand, we have noticed experimentally achieved
entanglement of six photons \cite{pan}, eight trapped ions \cite{eightion} and
twelve NMR ensemble qubits \cite{nmr-qubit}. So we also hope our idea could be
extended to other systems, such as to trapped ions. Furthermore, from above
discussion, we have known the experimental difficulty for ten atomic qubits to
be entangled and used for Grover search by current cavity QED technique.
Nevertheless, we argue that our study would be useful for future QIP
experiments by microwave cavity.

One of the authors (H.Q.F) would like to thank Professor Kelin Gao, Zhenyu Xu
and Xiaodong He for helpful discussions. This work is partly supported by
National Natural Science Foundation of China under Grants No. 10474118, No.
60490280, and No. 10774161, and partly by the National Fundamental Research
Program of China under Grants No. 2005CB724502 and No. 2006CB921203.

Fig. 1. Schematic setup for finding the marked state $\left\vert e_{1}%
i_{2}g_{3}g_{4}i_{5}i_{6}i_{7}i_{8}i_{9}i_{10}\right\rangle $ in a ten-qubit
Grover search, where ten atoms are simultaneously sent through the cavity with
proper speed. The ten state-selective field ionization detectors $D_{1}$,
$\cdot\cdot\cdot,$ $D_{10}$ are settled at the end of the passage for checking
the states of the qubits. The operations $U^{(10)},$ $H^{\otimes10},$
$S_{x,3},$ and $S_{x,4}$ are defined in the text. The inset shows the atomic
level configuration, with bold lines for the states encoding qubits and the
arrows for the coupling between the atoms and the standing wave of the cavity.

Fig. 2. The variations of the fidelity and probability of success with respect
to both $\eta$ and $\mu$.

Fig. 3. The search probability of ten-qubit with respect to the iteration
number under the cavity decay rate (a) $\mu=0$, (b) $\mu=0.05,$ and (c)
$\mu=0.1,$ respectively.

Fig. 4. Infidelity of ten-qubit phase gate versus time delay in the case of
$\eta=10$, where $\mu=0.05$ and $\mu=0.1$ are plotted by solid and dashed
curves, respectively.

\newpage%

\begin{figure}
[ptb]
\begin{center}
\includegraphics[
height=4.6622in,
width=6.9116in
]%
{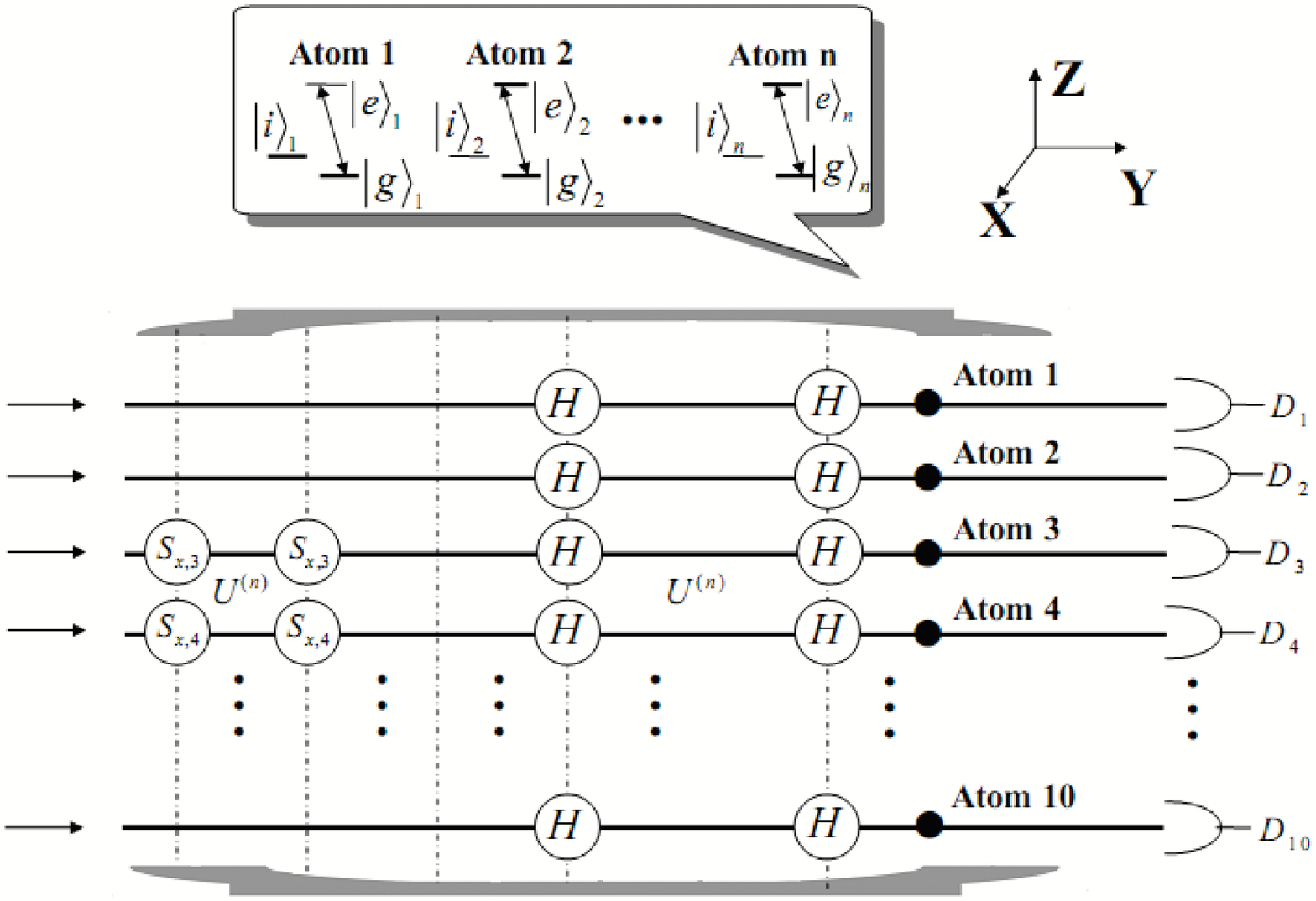}%
\end{center}
\end{figure}

\newpage%

\begin{figure}
[ptb]
\begin{center}
\includegraphics[
height=3.8527in,
width=8.073in
]%
{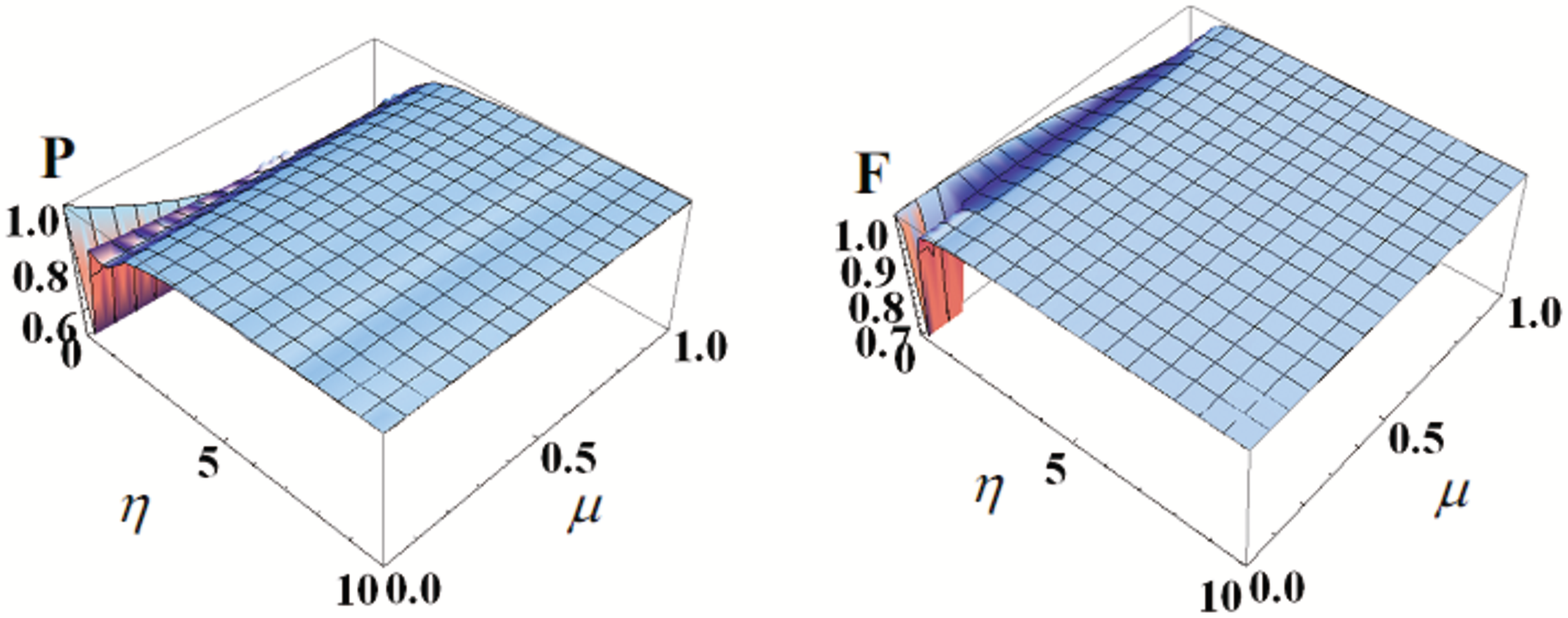}%
\end{center}
\end{figure}

\newpage%

\begin{figure}
[ptb]
\begin{center}
\includegraphics[
height=4.6639in,
width=6.1393in
]%
{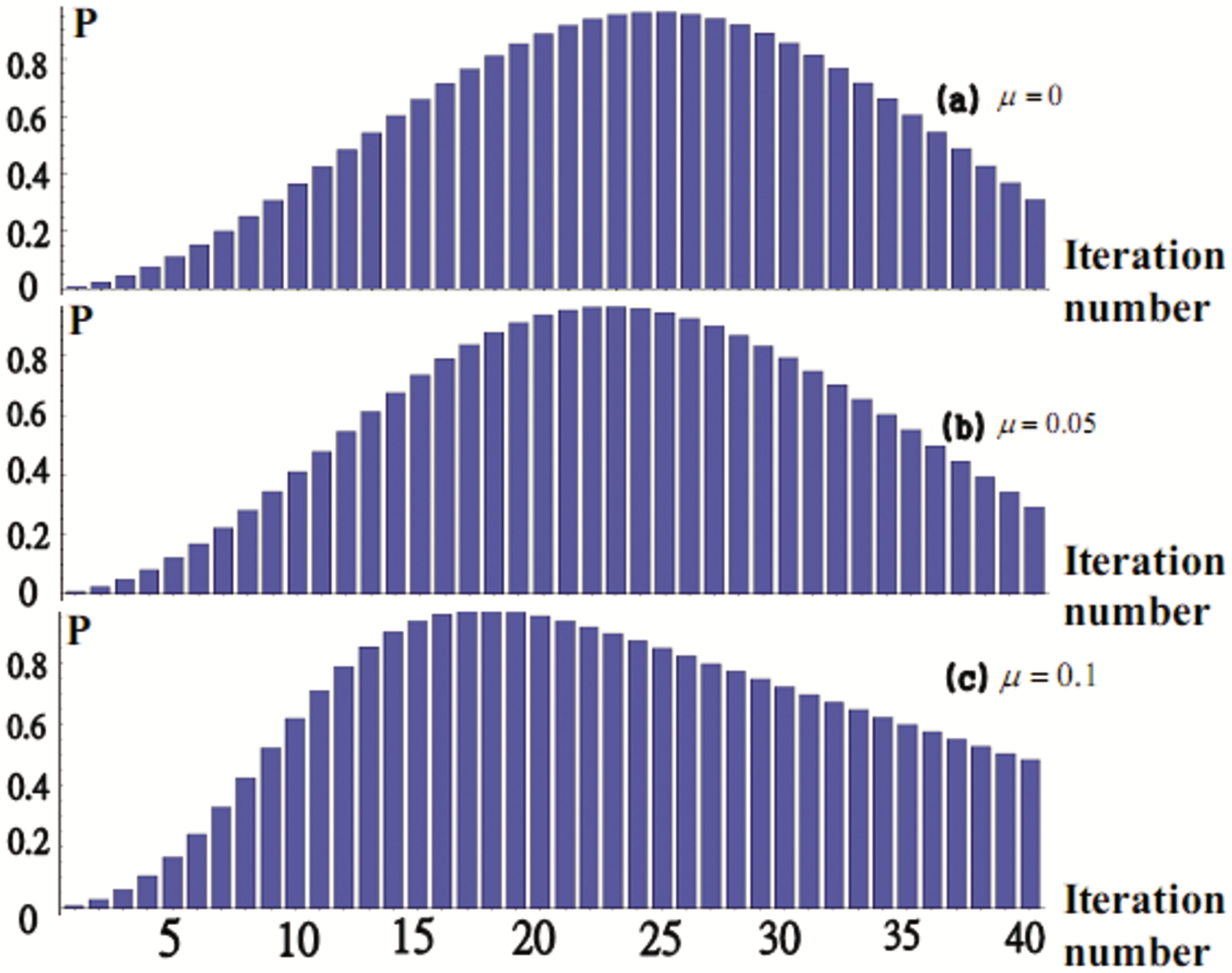}%
\end{center}
\end{figure}

\newpage%

\begin{figure}
[ptb]
\begin{center}
\includegraphics[
height=4.6613in,
width=6.4264in
]%
{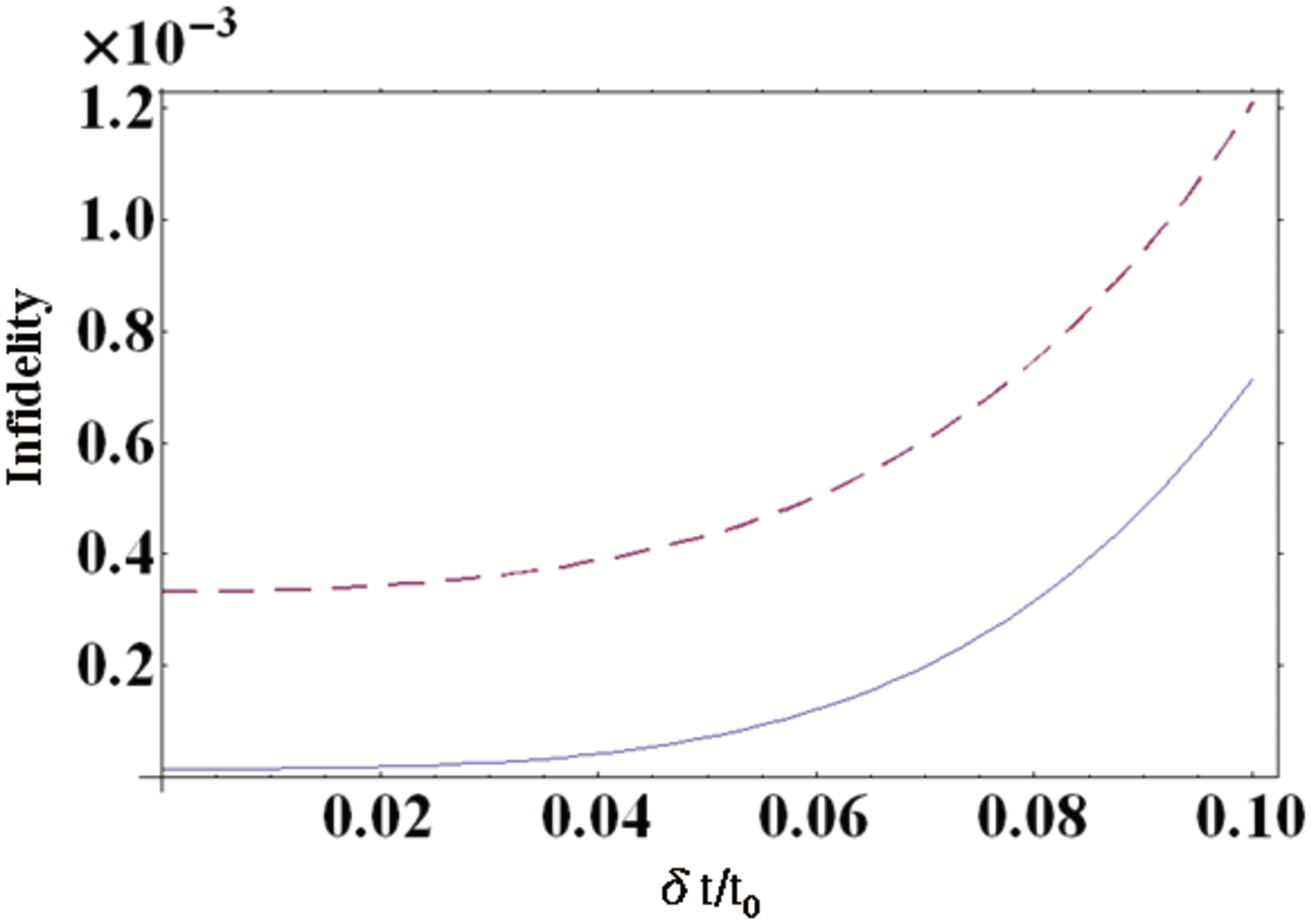}%
\end{center}
\end{figure}

\end{document}